\documentclass[aps,prl,reprint,superscriptaddress,amsmath,amssymb]{revtex4-2}
\usepackage[T1]{fontenc}
\usepackage{mathtools,placeins}
\usepackage{microtype}
\usepackage{slashed}
\usepackage{xcolor}
\usepackage{tikz,booktabs,graphicx,capt-of}
\usetikzlibrary{decorations.markings,decorations.pathmorphing}
\usepackage[hidelinks]{hyperref}
\allowdisplaybreaks[3]
\newcommand{\as}{a_s}
\newcommand{\Jg}{J_\gamma}
\newcommand{\Order}{\mathcal O}
\newcommand{\dd}{\mathrm d}
\newcommand{\euler}{\gamma_E}
\newcommand{\eps}{\epsilon}

\makeatletter
\newcommand{\lsection}[1]{\par\phantomsection\leavevmode\textit{#1.}---%
  \def\@currentlabel{\textit{#1}}\def\@currentlabelname{#1}\ignorespaces}
\newcommand{\lsubsection}[1]{%
  \par\leavevmode\phantomsection\textup{#1:}\quad
  \def\@currentlabel{\textit{#1}}\def\@currentlabelname{#1}\ignorespaces}

\newcommand{\emappendix}[2]{\par\setcounter{equation}{0}%
  \renewcommand{\theequation}{#1\arabic{equation}}%
  \renewcommand{\theHequation}{#1\arabic{equation}}%
  \leavevmode\phantomsection\textit{Appendix #1: #2}---%
  \def\@currentlabel{#1}\def\@currentlabelname{#2}\ignorespaces}
\makeatother

\begin{document}
\setcounter{secnumdepth}{3}
\title{Next-to-Next-to-Next-to-Leading-Order  Prediction for the Radiative \texorpdfstring{$B_{u}^- \to  \gamma \ell^{-} \bar \nu_{\ell}$}{Bu- to gamma ell- nu-bar} Decay}
\author{Yong-Kang Huang}
\affiliation{School of Physics, Nankai University, Weijin Road 94, Tianjin 300071, P.R. China}
\author{Yao Ji}
\affiliation{School of Science and Engineering, The Chinese University of Hong Kong, Shenzhen, Guangdong 518172, China}
\affiliation{Southern Center for Nuclear-Science Theory, Institute of Modern Physics, Chinese Academy of Sciences, Huizhou, Guangdong 516000, China}
\author{Bo-Xuan Shi}
\affiliation{Center for High Energy Physics, Peking University, Beijing 100871, China}
\author{Xing Wang}
\affiliation{School of Science and Engineering, The Chinese University of Hong Kong, Shenzhen, Guangdong 518172, China}
\affiliation{Southern Center for Nuclear-Science Theory, Institute of Modern Physics, Chinese Academy of Sciences, Huizhou, Guangdong 516000, China}
\author{Yu-Ming Wang}
\affiliation{School of Physics, Nankai University, Weijin Road 94, Tianjin 300071, P.R. China}
\author{Yu-Xiang Wei}
\affiliation{School of Science and Engineering, The Chinese University of Hong Kong, Shenzhen, Guangdong 518172, China}
\date{November 13, 2026}
\begin{abstract}

We accomplish for the first time the three-loop computation of  the photonic radiative jet function in soft-collinear effective theory, which constitutes the last missing and most difficult  ingredient  for the complete next-to-next-to-next-to-leading-order QCD  prediction of the simplest radiative $B_{u}^- \to  \gamma \ell^{-} \bar \nu_{\ell}$ decay with an energetic photon {at leading power}.
Importantly,  the most prominent  three-loop contribution to  the radiative bottom-meson decay form factors turns out to arise from the newly determined QCD correction to the jet function rather than from the recently obtained hard correction. 
We further demonstrate explicitly that our full three-loop calculation of these non-hadronic radiative $B$-meson decay form factors appear to be highly beneficial for an improved extraction of the fundamental  shape parameter $\lambda_B$ of  the leading-twist $B$-meson light-cone distribution amplitude.

\end{abstract}
\maketitle

\lsection{Introduction}
\label{sec:intro}
Radiative leptonic $B$-meson decay is sensitive to the light-cone
momentum distribution of the spectator quark. For an energetic photon, the leading-power form factors
factorize into a hard coefficient, a radiative jet and the leading-twist
$B$-meson light-cone distribution amplitude (LCDA).
{Its first inverse moment, $\lambda_B^{-1}$, is a key
nonperturbative parameter of the $B$ meson's light-cone structure
in the heavy-quark limit~\cite{Wang:2016qii,Beneke:2018wjp}.}
The radiative jet
function describes
photon emission from the spectator at the intermediate virtuality
$m_b\Lambda_{\rm QCD}$. {Partially integrated branching fractions provide access to the
inverse-moment parameter $\lambda_B$~\cite{Korchemsky:1999qb,DescotesGenon:2002mw,Lunghi:2002ju,Bosch:2003radiative,Beneke:2011nf,Braun:2012kp,Wang:2016qii,Wang:2018power,Beneke:2018wjp,Galda:2022rkn}.}

In soft-collinear effective theory
(SCET)~\cite{Bauer:2000yr,Bauer:2001yt,Beneke:2002ph}, the photonic
radiative jet function $J_\gamma$ arises from the interaction of a
soft quark with collinear quark and gauge fields in a subleading
Lagrangian~\cite{Beneke:2002ni}. In
$B^-\to\gamma\ell^-\bar\nu_\ell$, this interaction generates the
leading-power spectator amplitude~\cite{DescotesGenon:2002mw,Lunghi:2002ju,Bosch:2003radiative,Wang:2016qii}.
This amplitude involves a convolution of the radiative jet function ${J(E_\gamma\omega,\mu)=J_{\gamma}(-2E_\gamma\omega,\mu)}$ with the leading-twist LCDA $\phi_B^+(\omega,\mu)$. Here $E_\gamma$ is the photon energy, $\omega>0$ is the spectator
light-cone momentum and $\mu$ is the factorization scale.
{At tree level this convolution is fixed by the inverse moment
$\lambda_B^{-1}$. Radiative jet corrections also depend on logarithmic
LCDA moments, making the prediction sensitive to the LCDA shape at
fixed $\lambda_B${~\cite{Beneke:2018wjp,Galda:2022rkn}}.}

\begin{figure}[ht]
 \centering
 \begin{minipage}{0.38\columnwidth}
  \includegraphics[width=\linewidth]{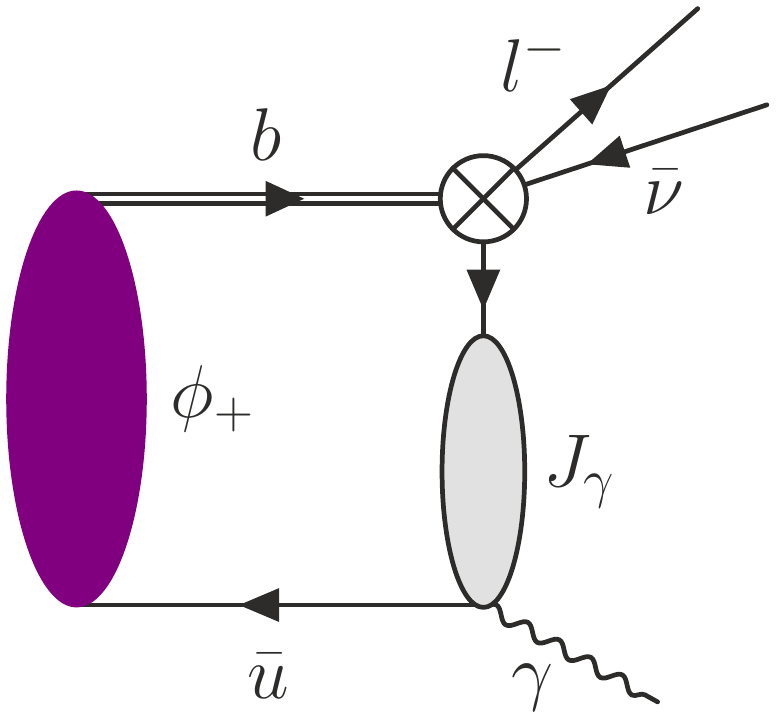}
 \end{minipage}
  \hspace{0.04\columnwidth}%
 \begin{minipage}{0.38\columnwidth}
  \includegraphics[width=\linewidth]{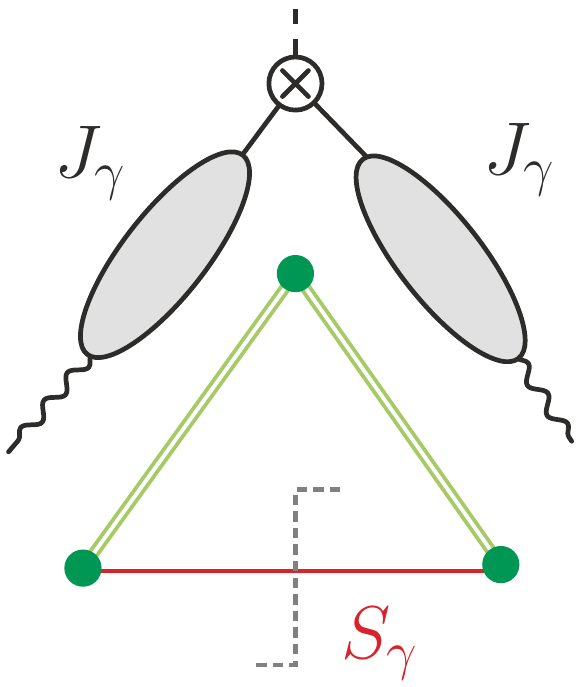}
 \end{minipage}
 \caption{The radiative jet function in $B^-\to\gamma\ell^-\bar\nu_\ell$ (left)
 and the light-quark-induced $h\to\gamma\gamma$ decay (right).}
 \label{fig:facts}
\end{figure}

The jet function was known through two loops~\cite{Liu:2020ydl}, while
the hard current and QCD--heavy-quark effective theory (HQET)
decay-constant matching have been calculated through three
loops~\cite{Fael:2024h2l,Bekavac:2009hqet}.
The three-loop jet function is the remaining short-distance matching
ingredient for a complete next-to-next-to-next-to-leading-order
(N$^3$LO) leading-power decay prediction. The two-loop analysis
found sizable corrections to the convolution at a hard-collinear
matching scale near 1.5 GeV, with their size depending on the
logarithmic moments~\cite{Liu:2020ydl,Galda:2022rkn}. The three-loop jet function contains
powers of $\ln(2E_\gamma\omega/\mu^2)$ up to degree six, so the
convolution probes logarithmic moments through sixth order.

{The evolution of the LCDA from a soft input scale to the jet
scale is governed by the Lange--Neubert (LN)
equation~\cite{Bell:2013dual,Galda:2020laplace,Galda:2022rkn}.
The one-loop kernel was first derived in Ref.~\cite{Lange:2003ff}.
Its two-loop correction was obtained using conformal
symmetry~\cite{Braun:2019wyx}.}
{Renormalization group (RG) consistency of the factorized
amplitude relates the jet anomalous dimension to the hard-current
anomalous dimension and the {LN kernel}.}
{We verify this RG consistency relation at three loops
using the companion LN kernel~\cite{Ji:LNcompanion}.}

We determine $J_\gamma$ at three loops with its full color and
electric-charge dependence using two independent approaches in
Feynman gauge. The first uses the SCET operator definition, and the
second extracts it from the hard-collinear region of the full-QCD
amplitude. The two calculations agree. Photon emission from a closed
quark loop introduces a new charge-weighted color structure at three
loops. Together with the hard matching, the result gives the complete
leading-power amplitude at N$^3$LO, and solving the hard and LN
evolution equations resums the logarithms between the hard, jet and
soft scales~\cite{Galda:2022rkn,Cui:2023B,Wang:2018power}.
We use the evolved form factor and normalization ratios to calibrate
$\lambda_B$.
{For an exponential $B$-meson LCDA, the new three-loop jet
coefficient lowers the convolution by about $4\%$.
The complete three-loop corrections also have a significant impact on
the extraction of $\lambda_B$.}

The same jet function occurs in the light-quark induced
$h\to\gamma\gamma$ form factor, shown in the right panel of
Fig.~\ref{fig:facts}. After refactorization-based subtraction (RBS), the
soft-quark-exchange term contains two radiative jets, one for each
photon direction, convolved with a soft-quark
function~\cite{Liu:2019oav,Liu:2020tzd,Liu:2020wbn,Beneke:2024soft}.
Our three-loop coefficient provides part of the boundary conditions
for resumming the large mass logarithms.

\lsection{Factorization and jet definition}
\label{sec:kinematics}
Consider the radiative leptonic decay
\begin{equation}
   B^-(p_B)\to\gamma(p_\gamma)\,\ell^-(p_\ell)\,\bar\nu_\ell(p_\nu)\,,
\end{equation}
with $\ell=e,\mu$
and negligible lepton masses. In the $B$-meson rest frame,
$p_B=m_Bv$ with $v=(n+\bar n)/2$, where $m_B$ is the meson mass and
$n$, $\bar n$ are lightlike vectors with $n^2=\bar n^2=0$ and
$n\cdot\bar n=2$. With the photon direction along $n$, its momentum
is $p_\gamma^\mu=E_\gamma n^\mu$, and the physical photon polarization
satisfies $n\cdot\varepsilon=\bar n\cdot\varepsilon=0$.
An arbitrary four-vector has the light-cone decomposition
$r^\mu=\frac{n\cdot r}{2}\bar n^\mu+\frac{\bar n\cdot r}{2}n^\mu
+r_\perp^\mu$. The transverse metric and Dirac matrices are
$g_\perp^{\mu\nu}=g^{\mu\nu}
-(n^\mu\bar n^\nu+\bar n^\mu n^\nu)/2$ and
$\gamma_\perp^\mu=g_\perp^{\mu\nu}\gamma_\nu$.
The soft momentum $k$ of the spectator antiquark $\bar u$ has
components of order $\Lambda_{\rm QCD}$. At leading power, the jet
function depends on it only through $\omega=n\cdot k>0$. We therefore
use the lightlike reference momentum $k^\mu=\omega\bar n^\mu/2$ in
the partonic matching calculation.
The momentum carried by the light-quark propagator between photon
emission and the weak interaction is then
$p^\mu=p_\gamma^\mu-k^\mu$,
with $p^2=-2E_\gamma\omega<0$.
We work at leading order in the electromagnetic interaction and at
leading power in the hierarchy
$m_b^2\sim E_\gamma^2\gg-p^2\sim E_\gamma\Lambda_{\rm QCD}
\gg\Lambda_{\rm QCD}^2$, where $m_b$ is the bottom-quark pole mass.

\lsubsection{Radiative jet operator}
The radiative jet function with an external photon originates from the
power-suppressed SCET interaction between a soft quark and
hard-collinear fields~\cite{Beneke:2002ph,Beneke:2002ni},
\begin{equation}
 \mathcal L_{q\xi_n}^{(1/2)}(x)
 =\bar q_s(x_-)W_n^\dagger(x)i\slashed D_{n\perp}\xi_n(x)
 +\mathrm{h.c.}\,,
 \label{eq:soft-collinear-Lagrangian}
\end{equation}
Here $q_s$ is the soft-quark field, evaluated at the
multipole-expanded coordinate $x_-^\mu=\frac{\bar n\cdot x}{2}n^\mu$, while $\xi_n$ is the collinear
quark field satisfying $\slashed n\xi_n=0$, and $W_n$ is the
collinear Wilson line along $\bar n$. The transverse covariant
derivative $D^\mu_{n\perp}$ contains both gluon and photon fields.
The superscript denotes suppression by
$(\Lambda_{\rm QCD}/m_b)^{1/2}$.

Factoring out the soft field in
Eq.~\eqref{eq:soft-collinear-Lagrangian}, with the collinear building
block $(\bar\xi_nW_n)(0)$ supplied by the weak current, the
bare radiative jet function is defined by the vacuum-to-photon matrix
element {in dimensional regularization
($D=4-2\epsilon$)}~\cite{Liu:2019oav,Liu:2020ydl,Liu:2021mac}:
\begin{align} \label{eq:photon-jet-operator}
 &\int\dd^D x\,e^{ip_s\cdot x_-}\langle\gamma(p_\gamma)|{\rm T}\!\left\{
 (W_n^\dagger i\slashed D_{n\perp}\xi_n)(x)
 (\bar\xi_nW_n)(0)\right\}|0\rangle\notag\\
 &={}e_q\slashed\varepsilon_\perp^*(p_\gamma)
 \frac{\slashed n}{2}\frac{i\bar n\cdot p_\gamma}{p^2+i0}\,
  J_\gamma^{\mathrm{bare}}(p^2),
 \end{align}
where $e_q=eQ_{\rm ext}$ is the electric charge of the open quark
line and $e>0$ is the positron charge.
The soft momentum, $p_s\propto\bar n$, flows out of the jet. Here,
only $p^2=(p_s+p_\gamma)^2=(\bar n\cdot p_\gamma)(n\cdot p_s)$ enters
the jet function. The prefactor in
Eq.~\eqref{eq:photon-jet-operator} is the tree-level matrix element,
so $J_\gamma^{(0)}=1$. For the incoming spectator in $B$ decay,
$p_s=-k$, giving the spacelike argument $p^2=-2E_\gamma\omega$.

\lsubsection{Hadronic inputs and factorization}
We denote the charged weak quark current by
$j_{\rm weak}^\mu=\bar u\gamma^\mu(1-\gamma_5)b$.
We define the vector and axial form factors through the transverse
hadronic tensor
\begin{align}
 &\int\!\dd^4x\,e^{ip_\gamma\cdot x}\langle0|{\rm T}\{j_{\nu,{\rm em}}^\perp(x)
      j_{\mu,{\rm weak}}^\perp(0)\}|B^-(p_B)\rangle_{\rm LP}
   \notag\\
 =&{} E_\gamma\bigl[g_{\mu\nu}^{\perp}F_A^{\rm LP}(E_\gamma)-i\epsilon_{\mu\nu\rho\sigma}\bar n^\rho v^\sigma
       F_V^{\rm LP}(E_\gamma)\bigr].
 \label{eq:transverse-form-factors}
\end{align}
The subscript LP denotes the leading-power term.
Here $j_{\nu,{\rm em}}=\sum_f Q_f\bar f\gamma_\nu f$ is the quark
electromagnetic current, and $Q_f$ is the charge of flavor $f$ in units
of the positron charge. A single transverse SCET operator determines
both structures at leading power, so $F_V^{\rm LP}=F_A^{\rm LP}$.
Their common value $F_{\rm LP}$ factorizes as~\cite{DescotesGenon:2002mw,Lunghi:2002ju,Bosch:2003radiative,Wang:2016qii}
\begin{align}
 {F_{\rm LP}}&{={}}
 \frac{Q_um_Bf_B}{2E_\gamma}\,{\mathcal I},\notag\\
 {\mathcal I}&{={}}H(E_\gamma,\mu)
 \int_0^\infty\!\frac{\dd\omega}{\omega}\,
 {J(E_\gamma\omega,\mu)}\phi_B^+(\omega,\mu).
 \label{eq:factorization-context}
\end{align}
Here $Q_u=2/3$ and ${J(E_\gamma\omega) = J_\gamma(-2E_\gamma\omega)}$. The physical decay constant $f_B$ and the static decay constant
$\widetilde f_B(\mu)$ are defined by
\begin{align}
 \langle0|\bar u\gamma^\rho\gamma_5 b|B^-(p_B)\rangle
 &=if_Bp_B^\rho,\notag\\
 \langle0|\bar u\gamma^\rho\gamma_5 h_v|B^-(p_B)\rangle
 &=im_B\widetilde f_B(\mu)v^\rho,
 \label{eq:hqet-decay-constant}
\end{align}
with the HQET field $h_v$, describing a static bottom quark with velocity $v$, and a relativistically normalized meson state. Matching the axial current at leading power gives
$f_B=K_F(m_b,\mu)\widetilde f_B(\mu)$, with the three-loop
conversion coefficient of Ref.~\cite{Bekavac:2009hqet}. $H=C_1/K_F$, where $C_1$ is the matching coefficient of
the transverse current from QCD to SCET, so that
$f_BH=\widetilde f_BC_1$. Wang and Shen~\cite{Wang:2018power} denote
$C_1$ by $C_\perp$ and use the jet function
{$J_\perp(2E_\gamma,\omega,\mu)={J(E_\gamma\omega,\mu)}$}. The leading-twist two-particle LCDA $\phi_B^+$ is defined by the
projected bilocal matrix
element~\cite{Grozin:1996pq,Lange:2003ff}
\begin{align}
 &\langle0|\bar u(zn)[zn,0]\,
       \slashed n\gamma_5 h_v(0)|B^-(p_B)\rangle
   \notag\\
 =&{}im_B\widetilde f_B(\mu)
   \int_0^\infty\!\dd\omega\,e^{-i\omega z}\phi_B^+(\omega,\mu).
 \label{eq:lcda-definition}
\end{align}
Here $z$ is the light-ray separation parameter, and $[zn,0]$ is the
straight soft Wilson line joining the quark fields.

\lsection{Three-loop calculation}
\label{sec:calculation}
We calculate the bare jet function at three loops using two independent
approaches in Feynman gauge as a cross-check.
\lsubsection{SCET matrix element}
After stripping the photon polarization vector from the left-hand
side of Eq.~\eqref{eq:photon-jet-operator}, we denote the remaining
Dirac matrix by $\mathcal M^\mu$. The bare jet function is extracted
with
\begin{equation}
	\label{eq:projector}
 J_\gamma^{\mathrm{bare}}(p^2)
 =\frac{p^2+i0}{i e_q\bar n\cdot p_\gamma}\,
  \frac{g_{\perp\mu\nu}}{2(D-2)}
  \operatorname{tr}_{D}\!\left[
  \mathcal M^\mu\frac{\slashed{\bar n}}{2}\gamma_\perp^\nu\right].
\end{equation}
Here $\operatorname{tr}_{D}$ denotes the Dirac trace, with the color unit
matrix factored out. We generate 1372 three-loop diagrams and the corresponding amplitudes
with \textsc{qgraf}~\cite{Nogueira:1991ex}, and then use
\textsc{Form}~\cite{Ruijl:2017dtg,Davies:2026cci} to perform the Dirac and
color algebra. We carry out these steps through interfaces provided by
\textsc{AmpRed}~\cite{Chen:2024xwt}, which also organizes the scalar
integrals into integral families. After integration-by-parts
reduction~\cite{Tkachov:1981wb,Chetyrkin:1981qh} with \textsc{Kira}~\cite{Klappert:2020nbg,Lange:2026kira3} and summing all contributions, 68 master integrals enter $J_\gamma$.

\lsubsection{Hard-collinear expansion in QCD}
We determine $\Jg$ from the full-QCD partonic current product in
Eq.~\eqref{eq:transverse-form-factors}, with the meson replaced by an
incoming on-shell bottom quark of momentum $p_b=m_bv$ and a soft antiquark
$\bar u(k)$. We generate the diagrams with
\textsc{FeynArts}~\cite{Hahn:2000kx} and perform the Dirac, Lorentz,
and color algebra in $D$ dimensions with
\textsc{FeynCalc}~\cite{Shtabovenko:2023idz}.
{We take all loop momenta to be hard-collinear and expand
to leading power in $\lambda$, where
$\lambda^2\sim\Lambda_{\rm QCD}/m_b$}~\cite{Beneke:1997zp}.
At leading power, the hard-collinear amplitude has the tree-level Dirac structure, and its scalar coefficient is $J_\gamma^{\rm bare}$.

A vector quark loop with one photon and zero or two gluons cancels under
charge conjugation, while the one-gluon color trace vanishes.
The first allowed photon-singlet contribution therefore appears at
{$\alpha_s^3$}, with one photon and three gluons attached to a massless quark
loop. Summing the massless loop flavors, with $Q_\Sigma=\sum_f Q_f$, gives
$Q_\Sigma/Q_{\rm ext}$ relative
to the tree amplitude, independently of the ordinary flavor multiplicity
$n_f$.  For $u,d,s$, $Q_\Sigma=0$. The 24 diagrams form four attachment classes, where $h=0,1,2,3$
counts the gluons attached to the heavy line. The six diagrams with
$h=3$ vanish at leading power, and the poles of the other three
classes cancel in their sum. The complete bare singlet sum is
therefore finite and independent of $p^2$ as $\epsilon\to0$.

After partial fractions and loop-momentum shifts, we reduce the nonsinglet integrals with \textsc{Kira}~\cite{Klappert:2020nbg,Lange:2026kira3}
and the singlet integrals with \textsc{FIRE}~\cite{Smirnov:2020fire},
using integration-by-parts identities~\cite{Tkachov:1981wb,Chetyrkin:1981qh}.

\lsubsection{Master integrals and evaluation}
\label{sec:integral-reduction}
{The master integrals depend on the single scale $s=p^2$.} Of the 68 masters, eight reduce to gamma functions and fourteen
are known form-factor integrals~\cite{Lee:2010ik,Lee:2010ug,Gehrmann:2005pd}.
We use two independent methods to calculate the remaining 46
masters as a cross-check. First, we evaluate them with
\textsc{AMFlow} 2.0~\cite{Liu:2018qbb,Liu:2020kpc,Liu:2021wks,
Liu:2022mfb,Liu:2022chg,Huang:2026amf} to 80 digits for $s=-1$ and
use the PSLQ algorithm~\cite{Ferguson:1999pslq} to fit the
coefficients through the required orders in $\epsilon$. The fitted
results are checked against numerical evaluations at $s=-4/5$.
Second, we calculate these masters analytically using the methods
described in Appendix~\ref{app:masters}.

\lsection{Renormalization and three-loop results}
\label{sec:setup}
The bare result is organized in terms of
the bare coupling $a_{s,0}=\alpha_{s,0}/(4\pi)$, with the
$\overline{\mathrm{MS}}$ factors absorbed into the integration
measure:
\begin{equation}
 J_\gamma^{\rm bare}(p^2)
 =1+\sum_{\ell=1}^3
 \left(\frac{a_{s,0}}{\mu^{2\epsilon}}\right)^\ell
 K_\ell(\epsilon,L)+\Order(a_{s,0}^4),
 \label{eq:bare-amplitude-expansion}
\end{equation}
with $K_\ell(\epsilon,L)=e^{-\ell\epsilon L}k_\ell(\epsilon)$ and
$L=\ln[(-p^2-i0)/\mu^2]$.
Appendix~\ref{app:bare-amplitudes} gives the Laurent coefficients $k_\ell$.

\lsubsection{Nonlocal jet renormalization}
{With $\as=\alpha_s(\mu)/(4\pi)$, coupling renormalization gives}
$a_{s,0}=\mu^{2\epsilon}\as Z_\alpha$, where $Z_\alpha$ is the
coupling counterterm. The jet function renormalizes nonlocally
as~\cite{Bosch:2003radiative,Liu:2020ydl}:
\begin{equation}
 J_\gamma(p^2,\mu)
 =\int_0^\infty\!\dd x\,
 Z_J(p^2,xp^2)J_\gamma^{\rm bare}(xp^2)\equiv Z_J\otimes J_\gamma^{\rm bare}.
 \label{eq:ren-convention}
\end{equation}
Here $x>0$ rescales the jet virtuality, and $Z_J$ admits the
perturbative expansion $Z_J=\sum_{n\geq0}\as^nZ_J^{(n)}$ with
tree-level normalization $Z_J^{(0)}=\delta(1-x)$.
Expanding $J_\gamma=\sum_{n\geq0}\as^nJ_\gamma^{(n)}(L)$ gives
\begin{align}
 &J_\gamma^{(3)}=\lim_{\epsilon\to0}\biggl[
 K_3-\frac{2\beta_0}{\epsilon}K_2
 +\biggl(\frac{\beta_0^2}{\epsilon^2}
        -\frac{\beta_1}{2\epsilon}\biggr)K_1\notag\\
 &\;+Z_J^{(1)}\otimes\Bigl(K_2-\frac{\beta_0}{\epsilon}K_1\Bigr)
 +Z_J^{(2)}\otimes K_1+Z_J^{(3)}\otimes1\biggr],
 \label{eq:finite-subtraction}
\end{align}
where $\beta_0=\frac{11}{3}C_A-\frac{4}{3}T_Fn_f$ and
$\beta_1=\frac{34}{3}C_A^2-\frac{20}{3}C_AT_Fn_f-4C_FT_Fn_f$ are the
coefficients of the QCD beta function.
The counterterm $Z_J^{(3)}\otimes1$ contains only poles.
The finite three-loop term requires $K_1$ through $\epsilon^4$,
$K_2$ through $\epsilon^2$, and $K_3$ through $\epsilon^0$.

The renormalization kernel $Z_J$ is determined by the jet anomalous
dimension, $\dd J_\gamma/\dd\ln\mu=-\gamma_J\otimes J_\gamma$.
Under $p^2=-2E_\gamma\omega$, scale independence of
Eq.~\eqref{eq:factorization-context}
gives~\cite{Liu:2020ydl,Galda:2022rkn,Ji:LNcompanion}
\begin{equation}
 \gamma_J=\left[\Gamma_{\rm cusp}\ln\frac{2E_\gamma}{\mu}
                   +\gamma_C\right]\delta(1-x)-\mathcal H_{\rm LN}^T.
 \label{eq:jet-LN-consistency}
\end{equation}
Here $\Gamma_{\rm cusp}$ is the fundamental lightlike cusp anomalous
dimension and $\gamma_C$ is defined by
$\dd\ln C_1/\dd\ln\mu=\Gamma_{\rm cusp}\ln(2E_\gamma/\mu)+\gamma_C$.
Our LN-kernel convention is
$\dd(\widetilde f_B\phi_B^+)/\dd\ln\mu
=-\mathcal H_{\rm LN}\otimes(\widetilde f_B\phi_B^+)$.
The transpose is taken with respect to $\dd\omega/\omega$.
We use the three-loop LN kernel constructed in the companion
paper~\cite{Ji:LNcompanion}.
{Appendix~\ref{app:jet-renormalization} gives the three-loop
counterterm $Z_J^{(3)}$.}
{All poles cancel in Eq.~\eqref{eq:finite-subtraction}.}

\lsubsection{Three-loop matching data}
At three loops, the coefficient contains six nonsinglet color
structures and a photon-singlet term:
\begin{align}\label{eq:color-decomposition}
 J_\gamma^{(3)}&=\frac{Q_\Sigma}{Q_{\rm ext}}J_\Sigma^{(3)}\\
 &\quad+C_F\big[C_F^2J_{FFF}+C_AC_FJ_{AFF}+C_A^2J_{AAF}\notag\\
 &\quad+C_FT_Fn_fJ_{FFf}+C_AT_Fn_fJ_{AFf}
 +T_F^2n_f^2J_{Fff}\big].\notag
\end{align}
Appendix~\ref{app:jet-coefficients} gives the full polynomials $J_i(L)$, whose highest power is $L^6$, and the known lower-order coefficients. The 21 nonzero coefficients of $L^2$ through $L^6$ follow from lower-loop matching and evolution data. The six linear logarithms also involve the three-loop LN kernel input. All 27 logarithmic coefficients agree with the evolution equation. The six values $J_i|_{L=0}$ supply the new nonsinglet matching constants. For fundamental quarks in $\mathrm{SU}(N)$ with $T_F=1/2$, the photon-singlet coefficient, with $\zeta_n$ the Riemann zeta values, is
\begin{equation}
 J_\Sigma^{(3)}=\frac{C_F(C_A^2-4)}{C_A}
 \Bigl(4+\frac{5\pi^2}{3}-\frac{\pi^4}{90}
 +\frac{14\zeta_3}{3}-\frac{80\zeta_5}{3}\Bigr).
 \label{eq:photon-singlet-coefficient}
\end{equation}
It is independent of $L$, and its color factor vanishes for
$\mathrm{SU}(2)$. For massless $u,d,s,c$, $n_f=4$ and $Q_\Sigma=2/3$.
The up-quark jet function in $B^-\to\gamma\ell^-\bar\nu_\ell$ has
$Q_{\rm ext}=Q_u=2/3$, giving $Q_\Sigma/Q_{\rm ext}=1$.

For $\mathrm{SU}(3)$ with these four light flavors,
the full jet function at the spacelike
matching point $L=0$ is, numerically,
\begin{equation}
 J_\gamma\big|_{L=0}\simeq1-3.52658\as-64.8606\as^2-1588.94\as^3.
 \label{eq:jet-matching-point}
\end{equation}
At $\mu=1.5$ GeV, our reference value
$\alpha_s^{(4)}(\mu)=0.35$ corresponds to $\as\simeq0.02785$.
The one-, two- and three-loop corrections are then $-9.8\%$,
$-5.0\%$ and $-3.4\%$ of the tree-level value, respectively.

\lsection{Radiative leptonic decay with three-loop matching}
\label{sec:phenomenology}
For the leading-power amplitude $\mathcal I$ in
Eq.~\eqref{eq:factorization-context}, we combine the three-loop jet
function with the three-loop heavy-to-light
current~\cite{Fael:2024h2l}, the decay-constant
matching~\cite{Bekavac:2009hqet}, and the companion LN kernel. The two-loop hard boundary is taken from
Ref.~\cite{Bell:2010currents}, and the finite three-loop current
is evaluated with FFh2l~\cite{Fael:2024h2l}.

\lsubsection{{RG-improved amplitude}}
The hard and soft boundary scales are $\mu_h$ and $\mu_s$, while
the convolution is evaluated at the jet scale $\mu_j$.
To resum the logarithms between these scales, we solve the
{hard-coefficient and LN RG equations}. With the
Laplace transform~\cite{Galda:2022rkn}
$\widetilde\phi_B^+(\eta,\mu)=\int_0^\infty(\dd\omega/\omega)\,
(\omega/\bar\omega)^{-\eta}\phi_B^+(\omega,\mu)$ with a {reference momentum}
$\bar\omega$,
the evolved amplitude takes the form
\begin{align}
 \mathcal I_{\rm RG}
 &=H(E_\gamma,\mu_h)U_H(E_\gamma,\mu_h,\mu_j)\notag\\
 &\times\left[
 J_\gamma(\partial_\eta,\mu_j){{\left(2E_\gamma\bar\omega/\mu_j^2\right)^\eta}}
 \widetilde\phi_B^+(-\eta,\mu_j)\right]_{\eta=0},
 \label{eq:pheno-resummed}
\end{align}

Here $J_\gamma(\partial_\eta,\mu_j)$ is the jet function with $L$
replaced by $\partial_\eta$, $U_H$ evolves $H=C_1/K_F$ from
$\mu_h$ to $\mu_j$, and $\widetilde\phi_B^+(-\eta,\mu_j)$ is evolved
from its boundary at $\mu_s$.

{We compare NLO+LL, NNLO+NLL and N$^3$LO+NNLL predictions.
Following Ref.~\cite{Ji:LNcompanion}, LL, NLL and NNLL denote evolution
with the complete LN kernels through one, two and three loops,
respectively.
{At matching order $\alpha_s^m$, with $m=1,2,3$,
hard and jet matching and local and nonlocal anomalous dimensions are
retained through $m$ loops. Cusp terms multiplying a scale logarithm
or $\partial_\eta$ are retained through $m+1$ loops, while the
nonlogarithmic cusp term in the LN evolution is retained through
$m$ loops. We use the four-loop cusp coefficient of
Ref.~\cite{Henn:2019swt}.}}
All orders use the same four-loop coupling with
$\alpha_s^{(4)}(1.5\,\mathrm{GeV})=0.35$.
{We use $(\mu_h,\mu_j,\mu_s)=(4.8,1.5,1.0)$ GeV
and four massless light flavors. The hard matching also includes
bottom-quark loops with pole mass $m_b=4.8$ GeV.}
Expanding Eq.~\eqref{eq:pheno-resummed}
in the common coupling reproduces the complete fixed-order amplitude
through $\as^3$.

\lsubsection{Moment coefficients and the jet contribution}
We organize the nonperturbative input in logarithmic moments
at $\mu_s$,
\begin{equation}
 \sigma_n=\lambda_B\int_0^\infty\frac{\dd\omega}{\omega}
            \ln^n\frac{\bar\omega}{\omega}\,
            \phi_B^+(\omega,\mu_s),
 \label{eq:pheno-moment-definitions}
\end{equation}
with $\lambda_B^{-1}=\int_0^\infty(\dd\omega/\omega)\,
\phi_B^+(\omega,\mu_s)$.
Choosing $\bar\omega$ such that $\sigma_1=0$ gives
\begin{equation}
 \mathcal I_{\rm RG}=
 \frac1{\lambda_B}\bigg[{\mathcal C_0}+\sum_{n\ge2}{\mathcal C_n}\sigma_n\bigg].
 \label{eq:pheno-moments}
\end{equation}
The coefficients ${\mathcal C_n}$ contain hard matching, jet matching and
evolution, while the moments $\sigma_n$ encode the nonperturbative
LCDA shape. At $E_\gamma=2.2$ GeV and $\bar\omega=0.3$ GeV,
our reference scales give ${\mathcal C_0=0.63129}$, ${\mathcal C_2=0.0368671}$,
${\mathcal C_3=0.0013879}$ and ${\mathcal C_4=0.0009663}$ at {N$^3$LO+NNLL}.
{At the same scales, NNLO+NLL gives
$\mathcal C_0=0.67765$.}

For the figures we use a boundary LCDA from the family
\begin{equation}
 \phi_B^+(\omega,\mu_s)
 =\frac{\omega^{\alpha-1}e^{-\omega/\omega_0}}
         {\Gamma(\alpha)\omega_0^\alpha},\qquad
 \omega_0=\frac{\lambda_B(\mu_s)}{\alpha-1},
 \label{eq:pheno-shape-family}
\end{equation}
where $\Gamma$ is Euler's gamma function and $\alpha>1$.
This family is a one-parameter specialization of the
models in Ref.~\cite{Beneke:2018wjp}. The main figures use the
exponential input $\alpha={2.0}$, for which $\sigma_1=0$ at
$\bar\omega=e^{-\gamma_E}\lambda_B$, where $\gamma_E$ is Euler's
constant.
We evolve its full Laplace transform.
At $\lambda_B({1.0}\,\mathrm{GeV})=350$ MeV and $E_\gamma=2.2$ GeV,
the {NLO+LL}, {NNLO+NLL} and {N$^3$LO+NNLL} form factors are
$F_{\rm LP}=({0.34005,0.31778,0.29962})$.
At fixed {NNLL} evolution, upgrading both matching
coefficients from two to three loops changes the form factor by
${-5.96}\%$.

To identify the jet contribution, we define
\begin{equation}
 R_J^{(n)}=
 \frac{\displaystyle\int_0^\infty\frac{\dd\omega}{\omega}\,
       {\sum_{k=0}^n\as^k{J^{(k)}(E_\gamma\omega,\mu_j)}}
       \phi_B^+(\omega,\mu_j)}
      {\displaystyle\int_0^\infty\frac{\dd\omega}{\omega}\,
       \phi_B^+(\omega,\mu_j)},
 \label{eq:pheno-jet-ratio}
\end{equation}
All ratios use the same three-loop LCDA evolution, so each step adds
a finite jet matching term.

\begin{figure}[!htbp]
\centering
\includegraphics[width=\columnwidth]{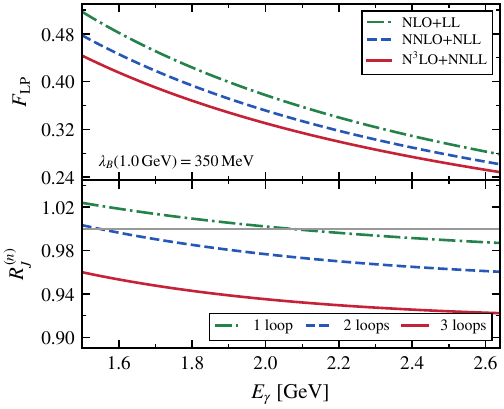}
\caption{RG-improved leading-power form factor (upper panel) and
successive jet matching contributions through the ratios
$R_J^{(n)}$ in Eq.~\eqref{eq:pheno-jet-ratio} (lower panel).
The boundary LCDA is exponential with
$\lambda_B(1.0\,\mathrm{GeV})=350$ MeV. The curves use the reference
scales.}
\label{fig:pheno-complete}
\end{figure}

{At $E_\gamma={2.0}$ GeV (lower panel of
Fig.~\ref{fig:pheno-complete}), the new three-loop jet coefficient
changes the ratio from $R_J^{(2)}=0.977$ to $R_J^{(3)}=0.935$.
The size and sign of this correction depend on the boundary LCDA shape.}
{At the same energy and fixed inverse moment, the
three-loop jet term changes the form factor by $-4.2\%$, $+2.5\%$
and $-5.8\%$ of the corresponding {NNLO+NLL} value for
$\alpha={2.0},1.7,2.5$, respectively.}

\lsubsection{Normalization ratios and the inverse moment}
To connect the matching correction to measurable quantities, we use
\begin{equation}
 R_\pi=\frac{\Delta\mathcal B}{\mathcal B(B^+\to\pi^0\ell^+\nu)},
 \quad
 R_\tau=\frac{\Delta\mathcal B}{\mathcal B(B^+\to\tau^+\nu)},
 \label{eq:pheno-ratios}
\end{equation}
where $\Delta\mathcal B=\tau_B\int_{E_{\rm cut}}^{m_B/2}
\dd E_\gamma\,\dd\Gamma/\dd E_\gamma$ is the partial
$B^+\to\gamma\ell^+\nu$ branching fraction above the photon-energy
cut $E_{\rm cut}$ and $\tau_B$ is the $B^+$ lifetime.
The theory ratios cancel $|V_{ub}|^2$, where $V_{ub}$ is the relevant
Cabibbo--Kobayashi--Maskawa matrix element, and $R_\tau$ also cancels
$f_B^2$ when the denominator is evaluated in the Standard Model.
{Both ratios use the Belle measurement with its photon-energy
cut $E_{\rm cut}=1.0$ GeV~\cite{Belle:2018pheno}.
The $R_\tau$ reference also uses the HFLAV average of
$\mathcal B(B^+\to\tau^+\nu)$~\cite{HFLAV:2026rare} (see End Matter).}

\begin{figure}[!htbp]
\centering
\includegraphics[width=\columnwidth]{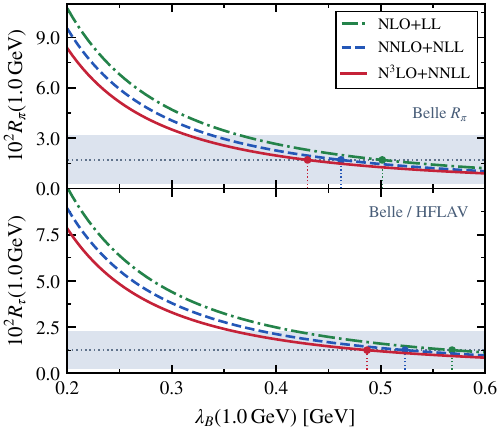}
\caption{RG-improved {leading-power} normalization ratios above a 1.0 GeV photon cut
as functions of $\lambda_B(1.0\,\mathrm{GeV})$ for the exponential
boundary LCDA. {The curves show the cumulative NLO+LL,
NNLO+NLL and N$^3$LO+NNLL predictions at the reference scales.}
Shading gives
the experimental one-standard-deviation reference intervals,
and the dots mark the central intersections. The lower interval
combines Belle's radiative measurement with the HFLAV average of
$\mathcal B(B^+\to\tau^+\nu)$.}
\label{fig:pheno-ratios}
\end{figure}

For the central $R_\pi$ measurement, the three curves in
Fig.~\ref{fig:pheno-ratios} select
$\lambda_B(1.0\,\mathrm{GeV})=({501.4,461.9,
429.9})$ MeV.
{At N$^3$LO+NNLL, this calibration is $32$ MeV lower
than at NNLO+NLL.}
The corresponding $R_\tau$ reference selects
$({568.3,523.2,487.0})$ MeV.
{With NNLL evolution and three-loop hard matching held fixed,
the new jet coefficient alone lowers the $R_\pi$ calibration
from $449.0$ to $429.9$ MeV.}
{For a narrower LCDA with $\alpha=2.5$ and
$\mu_j=1.2\,\mathrm{GeV}$, the N$^3$LO+NNLL calibration gives
$382$ MeV from $R_\pi$ and $436$ MeV from $R_\tau$.}

\lsection{Summary}
\label{sec:summary}
In conclusion, we have endeavored to carry out for the first time the three-loop computation of  the photonic radiative jet function in an analytical fashion  by applying the modern effective field theory formalism, enabling  us to achieve the full ${\rm N^3LO}$ QCD  prediction of the flagship  radiative $B_{u}^- \to  \gamma \ell^{-} \bar \nu_{\ell}$ decay at leading power accuracy.
It has been further demonstrated that the thus obtained three-loop QCD correction to the hard-collinear matching coefficient can bring about the substantial impact on the predicted radiative $B$-meson form factors with an energetic photon numerically.
In particular, taking  into account the complete three-loop QCD correction allowed for an improved determination of the  most significant  shape parameter of the leading-twist $B$-meson distribution amplitude.
Extending and developing further our three-loop  soft-collinear  factorization analysis to the  very rare  $B \to \gamma  \ell^{+}   \ell^{-} $ decay, to the four-body leptonic $B$-meson decay,
and to the exclusive electroweak penguin $B \to M \ell^{+} \ell^{-}$ decay will apparently be in high demand, on the one hand,  for exploiting the many facets of the strong interaction dynamics dictating exclusive heavy-hadron decay processes,  and on the other hand for probing sensitively the intriguing  non-standard quark flavour dynamics above the electroweak scale.

\vspace{0.5 cm}

\lsection{Note added}
\label{sec:Note Added}
The main results of this Letter  have already  been reported at an  international  workshop 
{\it Precision Across Scales: from QCD to the cosmos} [\url{https://indico.ph.tum.de/event/8036/}]
and   at a national   workshop on  cutting-edge research  of theoretical physics
[\url{https://indico.ihep.ac.cn/event/30082/}].

\vspace{0.5 cm}

\begin{acknowledgments}
\textit{Acknowledgments.} The  research of Y.K.H. was supported in part by  the China Postdoctoral Science Foundation and  Tianjin Joint  Support Program under Grant No. 2026T018TJ.
Y.J. is grateful for the support from  the National Natural Science Foundation of China with Grant No.12535006, from the Guangdong Basic and Applied Basic Research Foundation under Grant No. 2026A1515011306, and from the University Development Fund of the Chinese University of Hong Kong, Shenzhen, under the Grant No. UDF01003869.
{B.X.S. acknowledges support from the National Natural Science
Foundation of China under Grant No.~12325503.}
X.W. is supported by the University Development Fund of The Chinese University of Hong Kong, Shenzhen, under  Grant No. UDF01003912 and by the National Natural Science Foundation of China with Grant No. 12535006. Y.M.W. acknowledges support from the  National Natural Science Foundation of China with Grants No. 12475097 and No. 12535006, from  the Natural Science Foundation of Tianjin with Grant No. 25JCZDJC01190, and from the Fundamental Research Funds for the Central Universities with Grant No. 63261180. 
\end{acknowledgments}

\section*{End Matter}
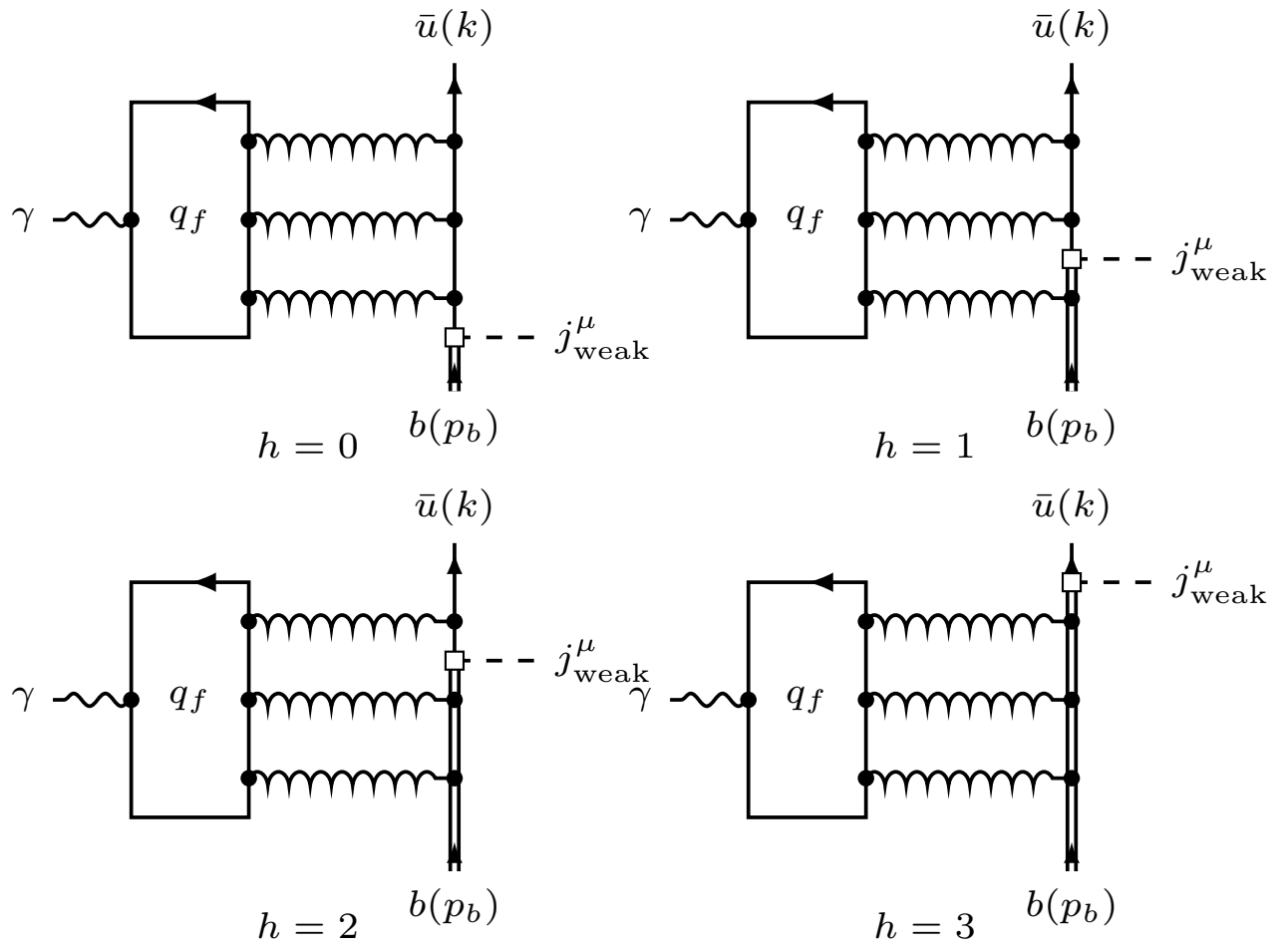
\begin{figure}[!htbp]
 \centering
 \begin{tikzpicture}[scale=0.65,font=\scriptsize,
   line/.style={line width=0.7pt},
   fermionloop/.style={line,postaction={decorate},
     decoration={markings,mark=at position 0.583333 with {\arrow{latex}}}},
   fermionline/.style={line},
   heavyline/.style={line,double,double distance=0.9pt},
   photon/.style={line,decorate,
     decoration={snake,amplitude=1.2pt,segment length=6pt}},
   gluon/.style={line,decorate,
     decoration={coil,amplitude=1.2pt,segment length=4pt}},
   vertex/.style={circle,fill=black,inner sep=1.1pt}]
  \foreach \h/\dx/\dy/\yw in
      {0/0/0/0,1/6.3/0/0.8,2/0/-4.9/1.6,3/6.3/-4.9/2.4}
  {\begin{scope}[shift={(\dx,\dy)}]
    \draw[fermionloop] (0,0) -- (1.2,0) -- (1.2,2.4)
        -- (0,2.4) -- cycle;
    \draw[photon] (0,1.2) -- (-0.8,1.2);
    \node[vertex] at (0,1.2) {};
    \node[left] at (-0.8,1.2) {$\gamma$};
    \node at (0.6,1.2) {$q_f$};
    \draw[heavyline] (3.3,-0.55) -- (3.3,\yw);
    \path[postaction={decorate},decoration={markings,
      mark=at position 0.5 with {\arrow{latex}}}]
        (3.3,-0.5) -- (3.3,0);
    \draw[fermionline] (3.3,\yw) -- (3.3,2.8);
    \path[postaction={decorate},decoration={markings,
      mark=at position 0.5 with {\arrow{latex}}}]
        (3.3,2.6) -- (3.3,2.8);
    \foreach \y in {2,1.2,0.4}
      {\draw[gluon] (1.2,\y) -- (3.3,\y);
       \node[vertex] at (1.2,\y) {};
       \node[vertex] at (3.3,\y) {};}
    \draw[line,dashed] (3.3,\yw) -- (4.15,\yw);
    \node[draw,fill=white,rectangle,inner sep=1.7pt] at (3.3,\yw) {};
    \node[right] at (4.15,\yw) {$j_{\rm weak}^\mu$};
    \node[above] at (3.3,2.8) {$\bar u(k)$};
    \node[below] at (3.3,-0.55) {$b(p_b)$};
    \node at (1.8,-1.1) {$h=\h$};
   \end{scope}}
 \end{tikzpicture}
 \caption{{Representative photon-singlet diagrams for the four
 attachment classes. The square marks the weak-current insertion
 $j_{\rm weak}^\mu$, and the double line denotes the heavy quark.
 The label $h$ counts gluons attached to the heavy-quark line.
 The $h=0,1,2$ classes contribute at leading power, while $h=3$
 is power suppressed.}}
 \label{fig:photon-singlet-topology}
\end{figure}

\emappendix{A}{{Photon-singlet contribution}}
{Photon emission from a closed massless quark loop first
contributes at $\alpha_s^3$. The three gluons connecting this loop
to the open quark line give the attachment classes in
Fig.~\ref{fig:photon-singlet-topology}. Each loop flavor is weighted
by its electric charge, giving $Q_\Sigma/Q_{\rm ext}$ relative to the
tree amplitude.}

{The sum over the two loop orientations produces the symmetric
color tensor $d^{abc}$, with
$d^{abc}d^{abc}/(2N)=C_F(C_A^2-4)/C_A$.
The complete leading-power contribution is finite and independent of
$L$, as given in Eq.~\eqref{eq:photon-singlet-coefficient}.}

\emappendix{B}{{Jet renormalization}}
\label{app:jet-renormalization}
{The LN kernel in Eq.~\eqref{eq:jet-LN-consistency} acts on the
HQET matrix element $\widetilde f_B\phi_B^+$.
With $\dd\ln\widetilde f_B/\dd\ln\mu=-\gamma_F$, the kernel for the
normalized LCDA is $\mathcal H_{\rm LN}-\gamma_F\mathbf1$.
The hard coefficient $H=C_1/K_F$ has the noncusp anomalous
dimension $\gamma_H=\gamma_C-\gamma_F$.
For a test function $f$, the transpose with respect to
$\dd\omega/\omega$ is
$(\mathcal H_{\rm LN}^Tf)(\omega')
=\omega'\int_0^\infty(\dd\omega/\omega)\,
\mathcal H_{\rm LN}(\omega,\omega',\mu)f(\omega)$.}

{Let $G_n$ and $\Gamma_n$ denote the coefficients of $\as^{n+1}$
in $\gamma_J$ and $\Gamma_{\rm cusp}$, respectively.
The operators $G_n$ follow from Eq.~\eqref{eq:jet-LN-consistency}.
Products act from right to left, and scalar terms multiply the identity.
The three-loop $\overline{\mathrm{MS}}$ counterterm is
\begin{align}
 Z_J^{(3)}={}&-\frac{\Gamma_0^3}{48\epsilon^6}
 +\frac{\Gamma_0^2(G_0-3\beta_0)}{16\epsilon^5}\notag\\
 &-\frac{\Gamma_0}{16\epsilon^4}
 \Bigl[G_0^2-5\beta_0G_0-\Gamma_1
                  +\frac{44}{9}\beta_0^2\Bigr]\notag\\
 &+\frac{G_0(G_0-2\beta_0)(G_0-4\beta_0)}
              {48\epsilon^3}\notag\\
 &-\frac{\Gamma_1G_0+2\Gamma_0G_1}{16\epsilon^3}
 +\frac{5\beta_0\Gamma_1+8\beta_1\Gamma_0}
              {36\epsilon^3}\notag\\
 &+\frac{G_0G_1+2G_1G_0}{24\epsilon^2}
 -\frac{3\beta_0G_1+3\beta_1G_0+\Gamma_2}
              {18\epsilon^2}\notag\\
 &+\frac{G_2}{6\epsilon}.
 \label{eq:jet-counterterm-three}
\end{align}
}

\emappendix{C}{{Numerical and experimental inputs}}
\label{app:numerical-inputs}
{For $F_V^{\rm LP}=F_A^{\rm LP}=F_{\rm LP}$, the photon-energy
spectrum is~\cite{Wang:2018power}
\begin{align}
 \frac{\dd\Gamma_{\rm LP}}{\dd E_\gamma}
 &=\frac{\alpha_{\rm em}G_F^2|V_{ub}|^2m_B}{3\pi^2}
 E_\gamma^3\notag\\
 &\quad\times\left(1-\frac{2E_\gamma}{m_B}\right)
 |F_{\rm LP}(E_\gamma)|^2.
 \label{eq:pheno-spectrum-LP}
\end{align}
The Standard-Model denominator of $R_\tau$ is evaluated from
\begin{equation}
 \Gamma(B^+\to\tau^+\nu)=
 \frac{G_F^2|V_{ub}|^2f_B^2m_Bm_\tau^2}{8\pi}
 \left(1-\frac{m_\tau^2}{m_B^2}\right)^2.
 \label{eq:pheno-tau-width}
\end{equation}
The common lifetime and $|V_{ub}|^2$ cancel in the theory ratios.
We use $m_B=5.279$ GeV, $f_B=0.190$ GeV,
$m_\tau=1.77686$ GeV, $\alpha_{\rm em}=1/137.035999084$ and
$G_F=1.1663787\times10^{-5}$ GeV$^{-2}$.}

Belle measures
$R_\pi=(1.7\pm1.4)\times10^{-2}$ and quotes
$\Gamma(B^+\to\pi^0\ell^+\nu)/|V_{ub}|^2=(2.4\pm0.2)\times10^{-12}$
GeV~\cite{Belle:2018pheno}.
For $R_\tau$, we combine Belle's partial radiative branching fraction
$(1.4\pm1.0\pm0.4)\times10^{-6}$ at the same photon cut with the
April 2026 HFLAV average~\cite{HFLAV:2026rare}
$\mathcal B(B^+\to\tau^+\nu)=(1.119\pm0.189)\times10^{-4}$.
Gaussian propagation of the quoted experimental uncertainties gives
the reference ratio
$R_\tau^{\rm ref}=({1.25\pm0.99})\times10^{-2}$.
{Both experimental references use the same Belle radiative sample.}

\clearpage
\onecolumngrid
\begin{center}{\Large\bfseries Supplemental Material\par}\end{center}
\appendix
\setcounter{figure}{0}\setcounter{table}{0}
\renewcommand{\thesection}{S\arabic{section}}
\renewcommand{\theequation}{\thesection.\arabic{equation}}
\renewcommand{\theHequation}{\thesection.\arabic{equation}}
\renewcommand{\thefigure}{S\arabic{figure}}\renewcommand{\theHfigure}{S\arabic{figure}}
\renewcommand{\thetable}{S\arabic{table}}\renewcommand{\theHtable}{S\arabic{table}}
\section{Perturbative jet coefficients}
\label{app:bare-amplitudes}
\label{app:jet-coefficients}
We write $\zeta_n=\zeta(n)$ for Riemann zeta values and denote
Euler's constant by $\gamma_E$.
This appendix collects the bare Laurent series and the renormalized
momentum-space polynomials in the conventions of
Section~\ref{sec:setup}. The common scale dependence of each bare
coefficient is $K_\ell(\epsilon,L)=e^{-\ell\epsilon L}k_\ell(\epsilon)$.
The color decomposition of the finite three-loop term is given in
Eq.~\eqref{eq:color-decomposition}.
\subsection{Bare coefficients}
Closed quark loops are treated as massless, and $n_f$ and $Q_\Sigma=\sum_f Q_f$ are independent flavor parameters.

\subsubsection{One loop}

The exact one-loop result, written in terms of Euler's gamma function,
is~\cite{Liu:2020ydl}
\begin{align}
 k_1(\epsilon)=C_F k_F^{(1)}(\epsilon)\,,\quad \text{with}\quad
 k_F^{(1)}(\epsilon)
 =e^{\euler\epsilon}(2-4\epsilon-\epsilon^2)
   \frac{\Gamma^2(-\epsilon)\Gamma(1+\epsilon)}
        {\Gamma(2-2\epsilon)}.
 \label{eq:bare-one-loop-exact}
\end{align}

\subsubsection{Two loops}

The two-loop color decomposition is
\begin{equation}
 k_2(\epsilon)=C_F\left[
   C_F k_{FF}^{(2)}(\epsilon)
  +C_A k_{AF}^{(2)}(\epsilon)
  +T_Fn_f k_{Ff}^{(2)}(\epsilon)\right].
 \label{eq:bare-two-loop-colors}
\end{equation}
Adding the photon and gluon contributions tabulated in Appendix~B of
Ref.~\cite{Liu:2020ydl} gives
\begingroup
\small
\begin{align}
&k_{FF}^{(2)}(\epsilon)={}\frac{2}{\epsilon^4}-\frac{1}{\epsilon^{2}}\Bigl(2 +\frac{\pi ^2}{3}\Bigr)-\frac{1}{\epsilon}\Bigl(4 +\frac{\pi ^2}{2} +\frac{46 \zeta _3}{3}\Bigr)-\frac{13}{2} -\frac{\pi ^2}{6} +\frac{\pi ^4}{5} -39 \zeta _3-\epsilon\Bigl(\frac{33}{4} -\frac{2 \pi ^2}{3} +\frac{11 \pi ^4}{15} +\frac{431 \zeta _3}{3} -\frac{26 \pi ^2 \zeta _3}{9} \notag\\*
&\qquad\quad{}-\frac{766 \zeta _5}{5}\Bigr)-\epsilon^{2}\Bigl(\frac{19}{8} -\frac{37 \pi ^2}{12} +\frac{116 \pi ^4}{45} -\frac{3707 \pi ^6}{5670} +\frac{1780 \zeta _3}{3} -\frac{89 \pi ^2 \zeta _3}{6} -\frac{2077 \zeta _3^2}{9} +153 \zeta _5\Bigr)+\Order(\epsilon^{3}).\label{eq:bare-2-FF}\\[2pt]
&k_{AF}^{(2)}(\epsilon)={}\frac{11}{6 \epsilon^3}+\frac{1}{\epsilon^{2}}\Bigl(\frac{67}{18} -\frac{\pi ^2}{6}\Bigr)+\frac{1}{\epsilon}\Bigl(\frac{103}{27} -\frac{11 \pi ^2}{36} -7 \zeta _3\Bigr)-\frac{695}{162} -\frac{103 \pi ^2}{108} \notag\\*
&\qquad{}-\frac{43 \pi ^4}{180} -\frac{14 \zeta _3}{9}-\epsilon\Bigl(\frac{42551}{972} +\frac{265 \pi ^2}{162} -\frac{31 \pi ^4}{720} -\frac{1142 \zeta _3}{27} -\frac{53 \pi ^2 \zeta _3}{18} +131 \zeta _5\Bigr)\notag\\*
&\qquad{}-\epsilon^{2}\Bigl(\frac{1115231}{5832} +\frac{1573 \pi ^2}{972} -\frac{319 \pi ^4}{432} +\frac{15311 \pi ^6}{45360} -\frac{22462 \zeta _3}{81} +\frac{47 \pi ^2 \zeta _3}{27} +\frac{64 \zeta _3^2}{3} +\frac{416 \zeta _5}{15}\Bigr)+\Order(\epsilon^{3}).\label{eq:bare-2-AF}\\[2pt]
&k_{Ff}^{(2)}(\epsilon)={}-\frac{2}{3 \epsilon^3}-\frac{10}{9 \epsilon^2}-\frac{1}{\epsilon}\Bigl(\frac{20}{27} -\frac{\pi ^2}{9}\Bigr)+\frac{230}{81} +\frac{5 \pi ^2}{27} +\frac{64 \zeta _3}{9}+\epsilon\Bigl(\frac{4591}{243} +\frac{10 \pi ^2}{81} +\frac{19 \pi ^4}{180} +\frac{320 \zeta _3}{27}\Bigr)\notag\\*
&\qquad{}+\epsilon^{2}\Bigl(\frac{113863}{1458} -\frac{115 \pi ^2}{243} +\frac{19 \pi ^4}{108} +\frac{640 \zeta _3}{81} -\frac{32 \pi ^2 \zeta _3}{27} +\frac{544 \zeta _5}{15}\Bigr)+\Order(\epsilon^{3}).\label{eq:bare-2-Ff}
\end{align}
\endgroup

\subsubsection{Three loops}

The three-loop result decomposes as
\begin{align}
 k_3(\epsilon)&=C_F\Bigl[
   C_F^2 k_{FFF}^{(3)}(\epsilon)
  +C_A C_F k_{AFF}^{(3)}(\epsilon)
  +C_A^2 k_{AAF}^{(3)}(\epsilon)\notag\\*
 &\qquad+C_FT_Fn_f k_{FFf}^{(3)}(\epsilon)
  +C_AT_Fn_f k_{AFf}^{(3)}(\epsilon)
  +T_F^2n_f^2 k_{Fff}^{(3)}(\epsilon)\Bigr]
  +\frac{Q_\Sigma}{Q_{\rm ext}}k_\Sigma^{(3)}(\epsilon).
 \label{eq:bare-three-loop-colors}
\end{align}
The six nonsinglet coefficients are
\begingroup
\small
\begin{align}
&k_{FFF}^{(3)}(\epsilon)={}\frac{4}{3 \epsilon^6}-\frac{1}{\epsilon^{4}}\Bigl(2 +\frac{\pi ^2}{3}\Bigr)-\frac{1}{\epsilon^{3}}\Bigl(4 +\pi ^2 +\frac{80 \zeta _3}{3}\Bigr)-\frac{1}{\epsilon^{2}}\Bigl(5 +\frac{\pi ^2}{2} -\frac{35 \pi ^4}{72} +82 \zeta _3\Bigr)-\frac{1}{\epsilon}\Bigl(\frac{1}{2} -\frac{4 \pi ^2}{3} +\frac{263 \pi ^4}{180} +294 \zeta _3 -\frac{20 \pi ^2 \zeta _3}{9} \notag\\*
&\qquad\quad{}-\frac{3928 \zeta _5}{15}\Bigr)-\frac{235}{12} +5 \pi ^2 -\frac{1133 \pi ^4}{144} +\frac{48875 \pi ^6}{54432} -\frac{4180 \zeta _3}{3} +\frac{341 \pi ^2 \zeta _3}{6}+\frac{1450 \zeta _3^2}{3} +414 \zeta _5+\Order(\epsilon).\label{eq:bare-3-FFF}\\[2pt]
&k_{AFF}^{(3)}(\epsilon)={}\frac{11}{3 \epsilon^5}+\frac{1}{\epsilon^{4}}\Bigl(\frac{67}{9} -\frac{\pi ^2}{3}\Bigr)+\frac{1}{\epsilon^{3}}\Bigl(\frac{313}{54} -\frac{11 \pi ^2}{12} -14 \zeta _3\Bigr)-\frac{1}{\epsilon^{2}}\Bigl(\frac{2587}{162} +\frac{173 \pi ^2}{36} +\frac{9 \pi ^4}{20} +41 \zeta _3\Bigr)\notag\\*
&\qquad{}-\frac{1}{\epsilon}\Bigl(\frac{48023}{486} +\frac{2069 \pi ^2}{216} -\frac{32993 \pi ^4}{12960} +256 \zeta _3 -\frac{25 \pi ^2 \zeta _3}{2} +\frac{826 \zeta _5}{3}\Bigr)-\frac{880025}{2916} -\frac{1675 \pi ^2}{648} +\frac{23729 \pi ^4}{7776} -\frac{190999 \pi ^6}{272160} -\frac{22397 \zeta _3}{18}\notag\\*
&\qquad{}+\frac{443 \pi ^2 \zeta _3}{108} +\frac{700 \zeta _3^2}{3} +\frac{45731 \zeta _5}{45}+\Order(\epsilon).\label{eq:bare-3-AFF}\\[2pt]
&k_{AAF}^{(3)}(\epsilon)={}\frac{242}{81 \epsilon^4}+\frac{1}{\epsilon^{3}}\Bigl(\frac{3254}{243} -\frac{44 \pi ^2}{81}\Bigr)+\frac{1}{\epsilon^{2}}\Bigl(\frac{7826}{243} -\frac{899 \pi ^2}{486} +\frac{22 \pi ^4}{405} -\frac{880 \zeta _3}{27}\Bigr)\notag\\*
&\qquad{}+\frac{1}{\epsilon}\Bigl(\frac{91997}{4374} -\frac{11641 \pi ^2}{1458} -\frac{253 \pi ^4}{162} -\frac{830 \zeta _3}{27} +\frac{44 \pi ^2 \zeta _3}{27} +32 \zeta _5\Bigr)-\frac{7551509}{26244} -\frac{250051 \pi ^2}{8748} -\frac{39959 \pi ^4}{19440} +\frac{5876 \pi ^6}{25515} +\frac{195623 \zeta _3}{486}\notag\\*
&\qquad{}+\frac{286 \pi ^2 \zeta _3}{9} +\frac{284 \zeta _3^2}{9} -\frac{10600 \zeta _5}{9}+\Order(\epsilon).\label{eq:bare-3-AAF}\\[2pt]
&k_{FFf}^{(3)}(\epsilon)={}-\frac{4}{3 \epsilon^5}-\frac{20}{9 \epsilon^4}-\frac{1}{\epsilon^{3}}\Bigl(\frac{34}{27} -\frac{\pi ^2}{3}\Bigr)+\frac{1}{\epsilon^{2}}\Bigl(\frac{328}{81} +\frac{13 \pi ^2}{9} +\frac{284 \zeta _3}{9}\Bigr)+\frac{1}{\epsilon}\Bigl(\frac{5255}{243} +\frac{133 \pi ^2}{54} -\frac{1891 \pi ^4}{3240} +\frac{4228 \zeta _3}{27}\Bigr)+\frac{111659}{1458} \notag\\*
&\qquad{}+\frac{109 \pi ^2}{81} +\frac{7609 \pi ^4}{9720} +\frac{61610 \zeta _3}{81} -\frac{205 \pi ^2 \zeta _3}{27} -\frac{2316 \zeta _5}{5}+\Order(\epsilon).\label{eq:bare-3-FFf}\\[2pt]
&k_{AFf}^{(3)}(\epsilon)={}-\frac{176}{81 \epsilon^4}-\frac{1}{\epsilon^{3}}\Bigl(\frac{2132}{243} -\frac{16 \pi ^2}{81}\Bigr)-\frac{1}{\epsilon^{2}}\Bigl(\frac{4156}{243} -\frac{212 \pi ^2}{243} -\frac{224 \zeta _3}{27}\Bigr)+\frac{1}{\epsilon}\Bigl(\frac{23240}{2187} +\frac{2659 \pi ^2}{729} +\frac{206 \pi ^4}{405} +\frac{2120 \zeta _3}{81}\Bigr)\notag\\*
&\qquad{}+\frac{1919567}{6561} +\frac{21623 \pi ^2}{2187} +\frac{1991 \pi ^4}{2430} -\frac{6736 \zeta _3}{81} -\frac{28 \pi ^2 \zeta _3}{3} +\frac{1232 \zeta _5}{3}+\Order(\epsilon).\label{eq:bare-3-AFf}\\[2pt]
&k_{Fff}^{(3)}(\epsilon)={}\frac{32}{81 \epsilon^4}+\frac{320}{243 \epsilon^3}+\frac{1}{\epsilon^{2}}\Bigl(\frac{16}{9} -\frac{8 \pi ^2}{81}\Bigr)-\frac{1}{\epsilon}\Bigl(\frac{12944}{2187} +\frac{80 \pi ^2}{243} +\frac{736 \zeta _3}{81}\Bigr)-\frac{415952}{6561} -\frac{4 \pi ^2}{9} -\frac{11 \pi ^4}{81} -\frac{7360 \zeta _3}{243}+\Order(\epsilon).\label{eq:bare-3-Fff}
\end{align}
\endgroup

For fundamental quarks in $\mathrm{SU}(N)$ with $T_F=1/2$, the complete
photon-singlet sum is
\begin{equation}
 k_\Sigma^{(3)}(\epsilon)
 =\frac{C_F(C_A^2-4)}{C_A}
  \left(4+\frac{5\pi^2}{3}-\frac{\pi^4}{90}
        +\frac{14\zeta_3}{3}-\frac{80\zeta_5}{3}\right)
  +\Order(\epsilon).
 \label{eq:bare-singlet-result}
\end{equation}
All its pole coefficients vanish after summing the attachment classes,
as detailed in the End Matter.

\subsection{Renormalized coefficients}
For convenience, we quote the tree-level, one-loop, and two-loop
coefficients from Refs.~\cite{Lunghi:2002ju,Bosch:2003radiative,Liu:2020ydl}:
\begin{equation}
 J_\gamma^{(0)}=1\,,\qquad
 J_\gamma^{(1)}(L)=C_F\left(L^2-1-\frac{\pi^2}{6}\right)\,,
 \label{eq:lo-nlo-momentum}
\end{equation}
and
\begin{equation}
 J_\gamma^{(2)}(L)=C_F\left[C_F J_{FF}(L)+C_A J_{AF}(L)+T_Fn_f J_{Ff}(L)\right],
 \label{eq:nnlo-momentum-decomposition}
\end{equation}
where
\begingroup
\small
\begin{align}
&J_{FF}(L)={}\frac{1}{2}L^{4} -\Bigl(1 +\frac{\pi ^2}{6}\Bigr)L^{2} +\Bigl(\pi ^2 +4 \zeta _3\Bigr)L +\frac{3}{2} -\frac{\pi ^2}{3} +\frac{119 \pi ^4}{360} -39 \zeta _3.\label{eq:jFF-momentum}\\[2pt]
&J_{AF}(L)={}-\frac{11}{9}L^{3} +\Bigl(\frac{67}{9} -\frac{\pi ^2}{3}\Bigr)L^{2} -\Bigl(\frac{305}{27} -14 \zeta _3\Bigr)L+\frac{493}{162} -\frac{103 \pi ^2}{108} -\frac{43 \pi ^4}{180} +\frac{140 \zeta _3}{9}.\label{eq:jAF-momentum}\\[2pt]
&J_{Ff}(L)={}\frac{4}{9}L^{3} -\frac{20}{9}L^{2} +\frac{76}{27}L +\frac{14}{81} +\frac{5 \pi ^2}{27} +\frac{8 \zeta _3}{9}.\label{eq:jFf-momentum}
\end{align}
\endgroup

The three-loop functions in Eq.~\eqref{eq:color-decomposition} are

\begingroup
\small
\begin{align}
&J_{FFF}(L)={}\frac{1}{6}L^{6} -\Bigl(\frac{1}{2} +\frac{\pi ^2}{12}\Bigr)L^{4} +\Bigl(\pi ^2 +\frac{28 \zeta _3}{3}\Bigr)L^{3} +\Bigl(\frac{3}{2} -\frac{\pi ^2}{3} +\frac{71 \pi ^4}{360} -51 \zeta _3\Bigr)L^{2} -\Bigl(\frac{\pi ^2}{2} +\frac{2 \pi ^4}{15} -44 \zeta _3 -\frac{32 \pi ^2 \zeta _3}{3} -28 \zeta _5\Bigr)L\notag\\*
&\qquad{}-\frac{281}{6} -\frac{3 \pi ^2}{2} -\frac{977 \pi ^4}{360} -\frac{55771 \pi ^6}{136080} -\frac{694 \zeta _3}{3} +\frac{95 \pi ^2 \zeta _3}{6} -\frac{290 \zeta _3^2}{3} +744 \zeta _5.\label{eq:jFFF-momentum}\\[2pt]
&J_{AFF}(L)={}-\frac{11}{9}L^{5} +\Bigl(\frac{67}{9} -\frac{\pi ^2}{3}\Bigr)L^{4} -\Bigl(\frac{272}{27} -\frac{11 \pi ^2}{54} -14 \zeta _3\Bigr)L^{3} -\Bigl(\frac{713}{162} +\frac{199 \pi ^2}{36} +\frac{11 \pi ^4}{60} -\frac{8 \zeta _3}{9}\Bigr)L^{2}\notag\\*
&\qquad{}+\Bigl(\frac{107}{27} +\frac{1129 \pi ^2}{81} -\frac{364 \pi ^4}{135} +\frac{2302 \zeta _3}{9} -\frac{29 \pi ^2 \zeta _3}{3} +40 \zeta _5\Bigr)L+\frac{5536}{81} -\frac{2395 \pi ^2}{972} +\frac{14015 \pi ^4}{1944} -\frac{2557 \pi ^6}{68040} -\frac{5437 \zeta _3}{9} -\frac{194 \pi ^2 \zeta _3}{27} \notag\\*
&\qquad{}+\frac{394 \zeta _3^2}{3} -\frac{1502 \zeta _5}{9}.\label{eq:jAFF-momentum}\\[2pt]
&J_{AAF}(L)={}\frac{121}{54}L^{4} -\Bigl(\frac{1780}{81} -\frac{22 \pi ^2}{27}\Bigr)L^{3} +\Bigl(\frac{13325}{162} -\frac{134 \pi ^2}{27} +\frac{11 \pi ^4}{45} -44 \zeta _3\Bigr)L^{2}-\Bigl(\frac{152795}{1458} -\frac{2192 \pi ^2}{243} -\frac{319 \pi ^4}{270} -\frac{3568 \zeta _3}{27} \notag\\*
&\qquad\quad{}+\frac{44 \pi ^2 \zeta _3}{9} +96 \zeta _5\Bigr)L-\frac{240323}{26244} -\frac{67655 \pi ^2}{4374} -\frac{31579 \pi ^4}{9720} +\frac{5876 \pi ^6}{25515} +\frac{57731 \zeta _3}{486} +\frac{275 \pi ^2 \zeta _3}{27}+\frac{284 \zeta _3^2}{9} -\frac{1954 \zeta _5}{9}.\label{eq:jAAF-momentum}\\[2pt]
&J_{FFf}(L)={}\frac{4}{9}L^{5} -\frac{20}{9}L^{4} +\Bigl(\frac{100}{27} -\frac{2 \pi ^2}{27}\Bigr)L^{3} -\Bigl(\frac{1291}{81} -\frac{17 \pi ^2}{9} -\frac{200 \zeta _3}{9}\Bigr)L^{2}+\Bigl(\frac{533}{9} -\frac{344 \pi ^2}{81} +\frac{92 \pi ^4}{135} -\frac{392 \zeta _3}{3}\Bigr)L-\frac{37357}{486} \notag\\*
&\qquad{}+\frac{1537 \pi ^2}{486} -\frac{3881 \pi ^4}{2430} +\frac{25052 \zeta _3}{81} +\frac{4 \pi ^2 \zeta _3}{27} -\frac{256 \zeta _5}{3}.\label{eq:jFFf-momentum}\\[2pt]
&J_{AFf}(L)={}-\frac{44}{27}L^{4} +\Bigl(\frac{1156}{81} -\frac{8 \pi ^2}{27}\Bigr)L^{3} -\Bigl(\frac{3310}{81} -\frac{40 \pi ^2}{27}\Bigr)L^{2} +\Bigl(\frac{11974}{729} -\frac{628 \pi ^2}{243} -\frac{34 \pi ^4}{135} -\frac{104 \zeta _3}{9}\Bigr)L+\frac{457499}{6561} +\frac{8321 \pi ^2}{2187} \notag\\*
&\qquad{}+\frac{194 \pi ^4}{243} -\frac{5900 \zeta _3}{81} -\frac{40 \pi ^2 \zeta _3}{27} +\frac{184 \zeta _5}{3}.\label{eq:jAFf-momentum}\\[2pt]
&J_{Fff}(L)={}\frac{8}{27}L^{4} -\frac{160}{81}L^{3} +\frac{256}{81}L^{2} +\Bigl(\frac{4496}{729} -\frac{64 \zeta _3}{27}\Bigr)L -\frac{132056}{6561} +\frac{8 \pi ^2}{243} +\frac{4 \pi ^4}{135} +\frac{320 \zeta _3}{243}.\label{eq:jFff-momentum}
\end{align}
\endgroup

\section{Master integrals}
\label{app:masters}
\nopagebreak[4]
\subsection{Normalization and integral families}
The 68 master integrals are normalized to be dimensionless:
\begin{equation}\label{eq:masterdef}
 M_{a, \boldsymbol{\nu}}
 =(-s)^{\nu-\frac{3D}{2}}\int\prod_{r=1}^{3}\frac{e^{\eps\gamma_E}\dd^D\ell_r}{i\pi^{D/2}}
 \prod_{j=1}^{12}\frac{1}{(-P_{a,j}-i0)^{\nu_j}}\,,
\end{equation}
Here $\boldsymbol{\nu}=(\nu_1,\ldots,\nu_{12})$ lists the propagator powers,
$\nu=\sum_{j=1}^{12}\nu_j$, and $\gamma_E$ is Euler's constant.
{The twelve propagator powers are written consecutively
in the compact labels used below.}
We use $s=p^2=(p_s+p_\gamma)^2=2p_s\cdot p_\gamma$, with $p_s^2=p_\gamma^2=0$.
The soft momentum $p_s$ and photon momentum $p_\gamma$ follow the conventions
of Section~\ref{sec:kinematics}. The masters are organized into the five
integral families in Table~\ref{tab:families}, with $a\in\{1,2,3,4,5\}$
labeling the family.
\begin{center}
\begin{minipage}{\linewidth}
\centering
\captionof{table}{The five integral families. Each denominator is $-P_{a,j}-i0$, with the
sign of $P_{a,j}$ fixed as shown.}
\label{tab:families}
\begingroup
\footnotesize\setlength{\tabcolsep}{4pt}
\begin{tabular}{@{}rlllll@{}}
\toprule
$j$ & $P_{1,j}$ & $P_{2,j}$ & $P_{3,j}$ & $P_{4,j}$ & $P_{5,j}$ \\
\midrule
$1$ & $\ell_3^2$ & $\ell_3^2$ & $\ell_3^2$ & $\ell_3^2$ & $\ell_3^2$ \\
$2$ & $\ell_2^2$ & $\ell_2^2$ & $\ell_2^2$ & $\ell_2^2$ & $\ell_2^2$ \\
$3$ & $\ell_1^2$ & $\ell_1^2$ & $\ell_1^2$ & $\ell_1^2$ & $\ell_1^2$ \\
$4$ & $(\ell_2+p_\gamma)^2$ & $(\ell_2+\ell_3)^2$ & $(\ell_2+p_\gamma)^2$ & $(\ell_2+p_\gamma)^2$ & $(\ell_2-p_s)^2$ \\
$5$ & $(\ell_1+\ell_2)^2$ & $(\ell_1-\ell_2)^2$ & $(\ell_2+\ell_3)^2$ & $(\ell_1+p_s)^2$ & $(\ell_1-p_\gamma)^2$ \\
$6$ & $(\ell_2-\ell_3+p_\gamma)^2$ & $(\ell_1+p_s+p_\gamma)^2$ & $(\ell_1-\ell_3)^2$ & $(\ell_1+\ell_2)^2$ & $(\ell_1+\ell_2)^2$ \\
$7$ & $(\ell_1+\ell_3+p_s)^2$ & $(\ell_1+\ell_3+p_s+p_\gamma)^2$ & $(\ell_1+p_s)^2$ & $(\ell_2-\ell_3-p_s)^2$ & $(\ell_1+\ell_3-p_\gamma)^2$ \\
$8$ & $p_s\!\cdot\!(\ell_3-\ell_2)-\tfrac{s}{2}$ & $-p_s\!\cdot\!(\ell_1+\ell_3)-\tfrac{s}{2}$ & $(\ell_1-\ell_2-\ell_3-p_\gamma)^2$ & $(\ell_2-\ell_3+p_\gamma)^2$ & $(\ell_1+\ell_3+p_s)^2$ \\
$9$ & $p_s\!\cdot\!(\ell_1+\ell_3)-\tfrac{s}{2}$ & $-p_s\!\cdot\!(\ell_2+\ell_3)-\tfrac{s}{2}$ & $(\ell_1-\ell_2-\ell_3+p_s)^2$ & $(\ell_1+\ell_3+p_s)^2$ & $(\ell_1+\ell_2+\ell_3)^2$ \\
$10$ & $(\ell_3-p_\gamma)^2$ & $(\ell_3+p_\gamma)^2$ & $p_s\!\cdot\!(\ell_1-\ell_2-\ell_3)-\tfrac{s}{2}$ & $p_s\!\cdot\!(\ell_3-\ell_2)-\tfrac{s}{2}$ & $p_s\!\cdot\!(\ell_1+\ell_3)-\tfrac{s}{2}$ \\
$11$ & $(\ell_1-p_\gamma)^2$ & $(\ell_1-\ell_2-\ell_3)^2$ & $(\ell_1-\ell_2-p_\gamma)^2$ & $(\ell_1-p_\gamma)^2$ & $(\ell_2+p_\gamma)^2$ \\
$12$ & $(\ell_1+p_s)^2$ & $(\ell_1-\ell_2+p_\gamma)^2$ & $-p_s\!\cdot\!\ell_2$ & $(\ell_1+\ell_2+p_s+p_\gamma)^2$ & $(\ell_1+p_s)^2$ \\
\bottomrule
\end{tabular}
\endgroup
\end{minipage}
\end{center}

Of the 68 masters, eight reduce to gamma functions and fourteen
are known form-factor integrals. The remaining 46 expansions are provided in
\texttt{RadJet3L-MasterIntegrals.m}, with their rational prefactors and
remainder orders. They use Eq.~\eqref{eq:masterdef} and are real for $s<0$.

\subsection{Analytic evaluation}

We start from the Feynman-parameter representation of
Eq.~\eqref{eq:masterdef} at $s=-1$, using the Cheng--Wu
freedom~\cite{Cheng:1987ga} to fix one quadratic Feynman parameter to unity.
For each representation, $\mathcal U$ and $\mathcal F$ denote its first and
second Symanzik polynomials. The eikonal propagators in
Table~\ref{tab:families} are linear in the loop momenta and all contain
$p_s$, so their Feynman parameters do not enter $\mathcal U$. Since
$p_s^2=0$, $\mathcal F$ is jointly linear in those parameters. Integrating
them removes one variable per eikonal line. The original $i0$ prescription
is understood in the reduced parameter integrals below.

For a single eikonal line with parameter $x_e$, we define
$\mathcal A$ and $\mathcal B$ by
$\mathcal F=\mathcal A-x_e\mathcal B/2$, with both independent of $x_e$.
The remaining integrations fall into the four groups in Table~\ref{tab:master-methods}.

\begin{center}
\begin{minipage}{\linewidth}
\centering
\captionof{table}{Analytic methods for the 46 master integrals.}
\label{tab:master-methods}
\begingroup
\small\setlength{\tabcolsep}{5pt}
\begin{tabular}{@{}lp{0.72\textwidth}r@{}}
\toprule
Group & Steps after eikonal integration & Number \\
\midrule
I & One-loop subintegrations and beta integrals, followed by
Euler-type integrals or convergent harmonic sums & 11 \\
II & As in group I, followed by hyperlogarithms with letters other
than $0$ and $1$ & 2 \\
III & Changes of variables, endpoint subtractions and
\textsc{HyperInt} integration & 24 \\
IV & Dimension shifts to $6-2\epsilon$ and reduction to
quasi-finite integrals & 9 \\
\bottomrule
\end{tabular}
\endgroup
\end{minipage}
\end{center}

\subsubsection{Subloop integration and convergent sums}
As a group-I example, we consider $M_{1,110110110000}$. The loop momentum $\ell_1$ appears only in
$P_{1,5}$ and $P_{1,7}$, which form a massless bubble with momentum
$q=\ell_3-\ell_2+p_s$. Integrating this bubble produces an effective
propagator $(-q^2-i0)^{-\eps}$. The remaining two-loop integral contains $P_{1,1}$, $P_{1,2}$, $P_{1,4}$, the
eikonal propagator $P_{1,8}$, and $q^2$ raised to the power $\eps$. The factor
$1/\Gamma(\eps)$ from the latter cancels the $\Gamma(\eps)$ of the bubble.
Denoting the parameters of $P_{1,1}$, $P_{1,2}$, and $P_{1,4}$ by $x_1$,
$x_2$, and $x_4$, respectively, and fixing the parameter of the effective
propagator to one, we obtain after eikonal integration
\begin{equation}
 \begin{aligned}
 M_{1,110110110000}
 &=-2e^{3\eps\gamma_E}\,
 \frac{\Gamma(-1+3\eps)\Gamma(1-\eps)^2}{\Gamma(2-2\eps)}\int_0^\infty\dd x_1\,\dd x_2\,\dd x_4\,
 \frac{\mathcal U^{-2+4\eps}\mathcal A^{1-3\eps}}{\mathcal B}\,,
 \end{aligned}
\end{equation}
with $\mathcal U=x_1x_2+x_1x_4+x_1+x_2+x_4$, $\mathcal A=x_1x_4$, and
$\mathcal B=x_1x_2+x_1+x_2+x_4$. The change of variables
$x_1=w/[(1-w)(1-y)]$, $x_2=w(1-v)/[(1-w)y]$, and
$x_4=wv/[(1-w)y]$, with $w,v,y\in(0,1)$, leaves Euler's beta integral $B(-\eps,1-\eps)$. This gives
\begin{align}
 M_{1,110110110000}
 &=-2e^{3\eps\gamma_E}\,
 \frac{\Gamma(-1+3\eps)\Gamma(1-\eps)^2}{\Gamma(2-2\eps)} B(-\eps,1-\eps)\,I(\eps)\,,\notag\\*
 I(\eps)
 &=\int_0^1\dd w\int_0^1\dd v\,
 \frac{w^{1-2\eps}(1-w)^{-2\eps}v^{1-3\eps}}{1-wv}
=\sum_{n=0}^\infty\frac{B(n+2-2\eps,1-2\eps)}{n+2-3\eps}\,.
\end{align}
The integral $I$ is finite at $\eps=0$, with
$I(0)=\zeta_2-1$ and $I'(0)=9\zeta_3-7$.
Together with the prefactor, these values determine
the double and simple poles of this master. Since the
summand falls off as $n^{-2+2\eps}$, the sum can be expanded in $\eps$ term by
term, and the resulting sums over harmonic sums reduce to multiple zeta values.
Integrating the expanded integrand with \textsc{HyperInt}~\cite{Panzer:2014caa}
gives the same coefficients. Because the prefactor has a double pole, $I$ is
needed through $\eps^4$.

The other masters of group I are evaluated in the same way. After the
one-loop, eikonal, and Beta integrations, a convergent one- or two-fold integral
of Euler type remains, which we expand through convergent sums or integrate
directly. For the two masters of group II, some coefficients come out as
hyperlogarithms with letters other than $0$ and $1$.
For $M_{5,101111110100}$ and $M_{5,110011110101}$,
the hyperlogarithms agree numerically with the multiple-zeta expressions
to 72--75 digits.

\subsubsection{Endpoint subtractions and hyperlogarithms}
As a group-III example, we consider $M_{2,111110101100}$. It has seven quadratic lines and one eikonal
line, and none of its loop momenta can be integrated out separately.
We assign $x_1,\ldots,x_7$ to the quadratic lines
$P_{2,1},P_{2,2},P_{2,3},P_{2,4},P_{2,5},P_{2,7},P_{2,10}$, respectively.
With $x_4$ fixed to one, the projective measure is
$\Omega=\delta(1-x_4)\prod_{j=1}^{7}\dd x_j$, with all $x_j\geq0$.
Integrating the parameter of the eikonal line $P_{2,9}$ gives
\begin{equation}
 M_{2,111110101100}=-2e^{3\eps\gamma_E}\,\Gamma(1+3\eps)
 \int\Omega\,\frac{\mathcal U^{4\eps}}{\mathcal A^{1+3\eps}\,\mathcal B}
\end{equation}
The parameter $x_6$ enters $\mathcal A$ only as an overall factor, and
$\mathcal U$ and $\mathcal B$ linearly. After a positive rational rescaling,
its integration separates a beta-function term $B(-3\eps,1-\eps)$ from a
remainder finite at $\eps=0$. The remaining five variables are transformed
by positive rational substitutions so that every divergence becomes a pole
at an endpoint of a single variable. These substitutions keep the integrand
linearly reducible~\cite{Brown:2008um,Panzer:2014caa}, i.e., in a suitable order
every polynomial that appears, including those produced by earlier
integrations, is linear in the next integration variable. We subtract the
endpoint poles of three variables, whose powers are
$-1-3\eps$, $-1-2\eps$, and $-1-\eps$. Together with the pole of the
Beta function, these subtractions produce the leading $1/\eps^4$ pole. Nesting the three
subtractions gives eight absolutely convergent integrals. We expand their
integrands in $\eps$ and integrate one variable at a time with
\textsc{HyperInt}. All periods reduce to multiple zeta values. Its coefficients have uniform weight, with weight
$6-n$ in the coefficient of $1/\eps^n$.

The other masters of group III follow the same steps with their own changes of
variables. Some of them also use parametric integration-by-parts identities or
partial fractions of the integrand. For $M_{1,100111111011}$, the $\eps^0$ and $\eps^1$ coefficients contain
hyperlogarithms with the letters $\pm2$ and $-1/2$, which \textsc{HyperInt} does
not reduce to multiple zeta values. We evaluate them to 400 digits with
\textsc{GiNaC}~\cite{Bauer:2000cp,Vollinga:2004sn} and determine their
expressions in multiple zeta values with an integer-relation
search~\cite{Ferguson:1999pslq}. The relations are stable from 100 to 400
digits.

\subsubsection{Dimension shifts and quasi-finite integrals}
The masters of group IV have nine or ten propagators. As an example, we
consider the nonplanar ten-line master $M_{4,011111111101}$,
whose parametric integral in $4-2\eps$
dimensions has overlapping divergences. Tarasov's dimension-shift
relations~\cite{Tarasov:1996br} express it through integrals in $6-2\eps$
dimensions with raised propagator powers. Each of the 44 monomials of
$\mathcal U$ generates a shifted integral. We reduce these 44 integrals with
\textsc{Kira}~\cite{Maierhofer:2017gsa,Klappert:2020nbg} and
\textsc{FireFly}~\cite{Klappert:2019emp} to 20 auxiliary integrals. We choose
14 of them, including the ten-line integral with one squared propagator, to be
quasi-finite~\cite{vonManteuffel:2014qoa}, i.e., their parametric integrals
converge at $\eps=0$ once a Gamma-function prefactor is removed. The other six auxiliary integrals are
products of Gamma functions. The poles of the master then come from the
reduction coefficients, which are rational functions of $\eps$ with poles up to
$1/\eps^5$, and from the Gamma-function prefactors. These poles determine the
order in $\eps$ to which each auxiliary integral is needed, from $\eps^1$ for
the ten-line integral to $\eps^5$ for the simplest ones. Eight of the
quasi-finite integrals become convergent sums over ratios of Gamma functions
after Beta integrations. One of them is the form-factor integral $A_{7,4}$ in
$6-2\eps$ dimensions, for which we use the representation of
Ref.~\cite{Heinrich:2007at}. The remaining six, including the ten-line
integral, are integrated over their Feynman parameters with \textsc{HyperInt}.
As a check of the reduction, 390 integration-by-parts identities generated
directly in momentum space vanish with the reduction rules.

The other masters of group IV are obtained in the same way. Some of their
auxiliary integrals follow from known three-loop form-factor
masters~\cite{Lee:2010ik,Lee:2010ug} through dimension-shift relations. For $M_{3,111110111110}$ and $M_{4,101111111110}$,
we replace auxiliary integrals of a lower sector, which would be needed to high
order in $\eps$, by quasi-finite integrals of a larger sector, which are needed
to lower order. The $\eps^1$ coefficient of $M_{3,111111111100}$ uses
higher-order coefficients of five form-factor masters from
Ref.~\cite{Lee:2010ik}. Its exact agreement with the numerical fit
checks this combination of coefficients.

The direct numerical evaluations agree with the fitted coefficients to more
than 75 digits. The analytic multiple-zeta expressions agree exactly with
the fitted ones. The five coefficients retained as hyperlogarithms in
group II reproduce them numerically to 72--75 digits.

\FloatBarrier

\end{document}